\documentclass[aps,a4paper, preprint, superscriptaddress,preprintnumbers,floatfix,nofootinbib,amsmath,amssymb,table]{revtex4-1}
\usepackage{hyperref}
\usepackage{cleveref}
\usepackage{subfig}
\usepackage{braket}

\usepackage{amstext,amssymb}
\usepackage[section]{placeins}
\usepackage{amsmath}
\usepackage{graphicx}
\usepackage{xspace}
\usepackage{color}
\usepackage{float}
\usepackage{comment}
\usepackage{amstext,amssymb}
\usepackage{amsmath}
\usepackage[section]{placeins}
\usepackage{xspace}
\usepackage{units}
\usepackage{array}
\usepackage{amsmath,bm}
\usepackage{amssymb}
\usepackage{gensymb}
\usepackage{soul}
\usepackage{textcomp}
\usepackage{slashed} 
\usepackage{multirow}
\usepackage{hyperref}
\usepackage{mathrsfs}
\usepackage{enumitem}
\usepackage{graphicx}
\usepackage{comment}
\usepackage{epstopdf}
\usepackage{float}
\usepackage{units}
\usepackage{feynmp}
\usepackage{slashed}
\usepackage{amsmath,bm}
\usepackage[table]{xcolor}
\usepackage[normalem]{ulem}
\definecolor{antiquewhite}{rgb}{0.98, 0.92, 0.84}
\definecolor{aqua}{rgb}{0.0, 1.0, 1.0}
\definecolor{amethyst}{rgb}{0.6, 0.4, 0.8}
\definecolor{applegreen}{rgb}{0.55, 0.71, 0.0}
\definecolor{byzantine}{rgb}{0.74, 0.2, 0.64}
\definecolor{cadetgrey}{rgb}{0.57, 0.64, 0.69}
\definecolor{candypink}{rgb}{0.89, 0.44, 0.48}
\usepackage{lipsum}
\usepackage{cals}
\usepackage{slashed} 
\newcommand{\be}{\begin{equation}}
\newcommand{\ee}{\end{equation}}
\newcommand{\bea}{\begin{eqnarray}}
\newcommand{\eea}{\end{eqnarray}}

\def\s1{\hat s}

\makeatletter

    \def\CT@@do@color{%
      \global\let\CT@do@color\relax
            \@tempdima\wd\z@
            \advance\@tempdima\@tempdimb
            \advance\@tempdima\@tempdimc
    \advance\@tempdimb\tabcolsep
    \advance\@tempdimc\tabcolsep
    \advance\@tempdima2\tabcolsep
            \kern-\@tempdimb
            \leaders\vrule
                    \hskip\@tempdima\@plus  1fill
            \kern-\@tempdimc
            \hskip-\wd\z@ \@plus -1fill }
    \makeatother

\usepackage{slashed}
\definecolor{ashgrey}{rgb}{0.7, 0.75, 0.71}
\definecolor{aureolin}{rgb}{0.99, 0.93, 0.0}
\definecolor{babypink}{rgb}{0.96, 0.76, 0.76}
\definecolor{buff}{rgb}{0.94, 0.86, 0.51}
\definecolor{chamoisee}{rgb}{0.63, 0.47, 0.35}
\definecolor{chartreuse(web)}{rgb}{0.5, 1.0, 0.0}
\definecolor{citrine}{rgb}{0.89, 0.82, 0.04}
\definecolor{emerald}{rgb}{0.31, 0.78, 0.47}
\definecolor{fawn}{rgb}{0.9, 0.67, 0.44}
\definecolor{fulvous}{rgb}{0.86, 0.52, 0.0}

\newcommand{\nua}[1]{\ensuremath{\rlap{\kern-2.5pt\ensuremath{\overset{\scriptscriptstyle(-)}{\phantom{\nu}}}}{\ensuremath{{\nu}_{#1}}}}\xspace}

\begin{document}
\title{Effect of torsion in long-baseline neutrino oscillation experiments}
\author{Papia Panda}
\email{ppapia93@gmail.com}
\affiliation{School of Physics,  University of Hyderabad, Hyderabad - 500046,  India}
\author{Dinesh Kumar Singha }
\email{dinesh.sin.187@gmail.com}
\affiliation{School of Physics,  University of Hyderabad, Hyderabad - 500046,  India}
\author{Monojit Ghosh}
\email{mghosh@irb.hr}
\affiliation{Center of Excellence for Advanced Materials and Sensing Devices,
Ruder Bošković Institute, 10000 Zagreb, Croatia}
\author{Rukmani Mohanta}
\email{rmsp@uohyd.ac.in}
\affiliation{School of Physics,  University of Hyderabad, Hyderabad - 500046,  India}

\begin{abstract}
In this work, we investigate the effect of curved spacetime on neutrino oscillation. In a curved spacetime, the effect of curvature on fermionic fields is represented by spin connection. The spin connection consists of a non-universal “contorsion” part which is expressed in terms of vector and axial current density of fermions. The contraction of contorsion part with the tetrad fields, which connects the internal flat space metric and the spacetime metric, is called torsion. In a scenario where neutrino travels through background of fermionic matter at ordinary densities in a curved spacetime, the Hamiltonian of neutrino oscillation gets modified by the torsional coupling constants $\lambda_{21}^{\prime}$ and $\lambda_{31}^{\prime}$. The aim of this work is to study the effect of $\lambda_{21}^{\prime}$ and $\lambda_{31}^{\prime}$ in DUNE and P2SO. In our study we, (i) discuss the effect of torsional coupling constants on the neutrino oscillation probabilities, (ii) estimate the capability of P2SO and DUNE to put bounds on these parameters and (iii) show how the physics sensitivities get modified in presence of torsion.

\end{abstract}

\maketitle
\flushbottom

\section{Introduction }
Neutrinos are the most enigmatic particles of nature. 
The absence of the right-handed partners for neutrinos within the Standard Model (SM) of particle physics  prevents them from getting mass via the conventional Higgs mechanism.
Consequently,  neutrinos are considered to be massless within the SM framework. 
However, from the experimental perspective, it has been firmly established that
 neutrinos  oscillate from one flavor to another during their propagation in
 matter or even in a vacuum, which is possible only if they posses non-zero masses. This intriguing behaviour necessitates an extension beyond the SM or a more comprehensive consideration of all interactions neutrinos might undergo. 
 Various beyond the Standard model mechanisms have been proposed in the past to explain the non-zero neutrino mass, such as the seesaw mechanism \cite{Ma:2006km,Davidson:2008bu,Ma:1998dn,Dienes:1998sb,Schechter:1980gr,Mohapatra:2006gs,Babu:1989rb,Malinsky:2005bi}, adding extra dimensions \cite{Dvali:1999cn,Clifton:2011jh,Arkani-Hamed:1999ylh,Arkani-Hamed:2000hpr,Arkani-Hamed:2003xts,Mohapatra:1999zd,Pas:2005rb,Roy:2023dyq,Garbrecht:2020tng,DeGouvea:2001mz,Lukas:2000wn}, or super-symmetry theory \cite{Bertone:2004pz,Fayet:1977yc,Mohapatra:1986bd,Barbier:2004ez}. Another alternative and crucial approach would be to ensure that all known physics, including gravitational interactions — to which all fermions and bosons are subjected to, are considered in the neutrino oscillation calculations.

The presence of curved spacetime affects the characteristics of the fermions. In a flat space, the Dirac equation explains the spin angular momentum of fermions by combining quantum mechanics and the special theory of relativity. However, when considering fermions in curved spacetime, the traditional use of the Dirac equation needs to be amended due to the impact of spacetime curvature on fermion propagation. Therefore, it is essential to modify the  Dirac equation to account for curved spacetime, incorporating the effects of curvature on fermionic fields. This adaptation involves encompassing the spin connection \cite{Ashtekar:1987gu}, a mathematical concept that dictates the behavior of spinors in curved spacetime by considering both the geometric curvature of spacetime and the intrinsic spin of the particle. The spin connection consists of two terms; one is solely the gravitational part, and the other is the non-universal ``contorsion" part. The contraction of contorsion part with the tetrad fields \cite{Maluf:2013gaa,Oliveri:2019gvm,Harst:2012ni,Shirafuji:1997wy,Nissinen:2018dnq} gives the antisymmetric part of the Christoffel symbol ($\Gamma$), which is called torsion. By changing the definition of the covariant derivative for spinor fields, the modified covariant derivative is utilized in the curved spacetime version of the Dirac equation to describe the dynamics of fermions in a gravitational field accurately. The spin connection provides a coherent framework for explaining the behavior of fermions in the presence of gravity, bridging the gap between the microscopic realm of quantum particles and the macroscopic curvature of spacetime as described by general relativity.

In this work, we have investigated the effect of gravity in neutrino oscillation to have a complete formalism by considering all the known physics in nature. The presence of curved spacetime modifies the expression of appearance and disappearance probabilities, which subsequently affect all sectors of neutrino oscillation physics, i.e., determination of CP violation, mass ordering and the octant of the atmospheric mixing angle. Previously in Ref.~\cite{Barick:2023wxx}, the authors discussed the effect of torsion in the neutrino oscillation in terms of probability. In our work, we intend to address how the neutrino physics sensitivities get affected in the presence of new torsional couplings. Our findings suggest that the upcoming long-baseline neutrino oscillation experiments like DUNE \cite{DUNE:2021cuw} and P2SO \cite{Akindinov:2019flp, Hofestadt:2019whx} may provide valuable insights into the torsional coupling constants. These two experiments are expected to have the best sensitivity towards the measurement of the neutrino oscillation parameters because of their longer baselines and larger statistics. To the best of our knowledge, our work is the first of its kind to estimate the sensitivity of the neutrino oscillation experiments in presence of gravity.

The structure of this paper is organized as follows. In Sec. \ref{3nu}, we show how the neutrino oscillation Hamiltonian gets modified in the presence of torsion and illustrate the oscillation probability expressions in the presence of torsion. Section \ref{experi} provides the details of the experimental setup and simulation strategy for the upcoming long-baseline neutrino experiments, namely DUNE and P2SO. In section \ref{tor-prob}, we explore the impact of torsion on both the appearance and disappearance channels of neutrino oscillations. Section \ref{bound-tor} discusses the experimental constraints placed on the  torsional couplings. Following that, section \ref{phys} examines alterations in neutrino physics sensitivities, including CP violation, mass ordering, and octant sensitivity, as influenced by the torsional coupling constants. The paper concludes with section \ref{summary}, offering a recapitulation of the main findings.

\section{Neutrino  oscillation in presence of torsional couplings}
\label{3nu}

Extensive  work has been done in the literature on neutrino oscillation in the presence of curved spacetime, see e.g., Refs.~\cite{Cruceru:2018jfq,Sousa:2009sk,Mavromatos:2012cc,Capozziello:2013dja,Zubkov:2013zxa,Pasic:2014apa,Fabbri:2015jaa,deAndrade:2016pht,deAndrade:2016vcz,Chakrabarty:2018ybk,Lin:2023sgn,Sun:2024qis,Adak:2000tp,SenGupta:2001zz,Adak:2003qg,Ringwald:2004te,Alimohammadi:1999zb,DeSabbata:1981ek,Ghose:2023ttq,Chakrabarty:2019cau,Mavromatos:2013osa,Barick:2023wxx,Nicolescu:2013rxa,Mandal:2021dxk}. Motivated by the notable research works that have been carried out in this area,  we  adopt the Einstein-Cartan-Sciama-Kibble (ECSK) formalism in our work. In this section, we will discuss the effect of torsion in three flavor neutrino oscillation. In our analysis we adopt the formalism as given in Refs.~\cite{Barick:2023wxx,Chakrabarty:2019cau}. The details of the formalism we provide in Appendix \ref{append} while here we just show how curved spacetime can affect the Hamiltonian of the neutrino oscillation under this formalism. As mentioned in the introduction, in a curved spacetime, the effect of curvature on fermionic fields is represented by spin connection. The spin connection consists of a non-universal “contorsion” part. The contraction of contorsion part with the tetrad fields is called torsion. In a scenario where neutrino travels through background of fermionic matter at ordinary densities in a curved spacetime, the additional interaction term contributing  to the effective  Hamiltonian can be written as \cite{Chakrabarty:2019cau}
\begin{equation}\label{L-nu.eff}
 		 H_{\rm {tor}} = \left(\sum\limits_{f=e,p,n}\lambda_{f} n_f \right) \left(\sum_{i=1,2,3} \lambda_{i}^V \bar{\nu}_i \gamma^0 \nu_i + \lambda_i^A \bar{\nu}_i \gamma^0 \gamma^5 \nu_i \right) \,,
 \end{equation}
where $n_f$ is the  number density of the background fermion and 
$\lambda_{i}^V (\lambda_i^A)$ 
is the torsional coupling for the vector (axial vector) components of the neutrinos. In eq. \ref{L-nu.eff}, the total contribution is coming from both left and right-handed neutrinos, however, for simplicity, we consider negligible torsional coupling for right-handed neutrinos. So, the effective Hamiltonian can be expressed as \cite{Barick:2023wxx},
\begin{equation}\label{H-nu.eff}
 	H_{\rm {tor}} = \sum_{i=1,2,3}\left(\lambda_{i}{\nu}_i^\dagger \mathbb{P_L} \nu_i  \right)\,\tilde{n}\,,
 \end{equation}
 where $\tilde n= \sum\limits_{f=e,p,n}(\lambda_{f} n_f) $, $\mathbb{P_L}=(1-\gamma_5)/2$ and 
 the term $\lambda_i=(\lambda_i^V-\lambda_i^A)$
  includes both the vector and axial vector components.
Thus, the total Hamiltonian includes contributions from both the vector and axial vector currents. 
 Therefore, the complete Hamiltonian can be written as,
\begin{equation}
    H_{\rm total} = H_{\rm vac} + H_{\rm mat} + H_{\rm tor}\;,
\end{equation}
where $H_{\rm vac}$ and $H_{\rm mat}$ are the Hamiltonian in vacuum and matter respectively. The
additional term $(H_{\rm tor})$ couples to the mass eigenstates of neutrinos as curved spacetime affects the mass of neutrinos directly. The time evolution of mass eigenstates in presence of torsion is~\cite{Barick:2023wxx} 
\begin{align}
& i\frac{d}{dt}\begin{pmatrix}{\nu_1}\\ {\nu_2} \\ \nu_3\end{pmatrix}=\Biggr[\frac{1}{2E}\begin{pmatrix}m_1^2 & 0 & 0 \\ 0 & m_2^2 & 0 \\ 0 & 0 & m_3^2\end{pmatrix}+\begin{pmatrix}\lambda_1 & 0 & 0 \\ 0 & \lambda_2 & 0 \\ 0 & 0 & \lambda_3\end{pmatrix}\tilde{n} \nonumber\\
 &\quad\quad\quad\quad\quad
 +U^{T}\begin{pmatrix}\sqrt{2} G_F n_e & 0 & 0 \\ 0 & 0 & 0 \\ 0 & 0 & 0\end{pmatrix}U^{*}\Biggr]\begin{pmatrix}\nu_1 \\ \nu_2 \\ \nu_3 \end{pmatrix}\,, \label{eq:TDSE_for_3nu.a}
 \end{align}
where $n_e$ is the electron number densities respectively and $U$ is the PMNS mixing matrix containing three mixing angles: $\theta_{12}$, $\theta_{23}$ and $\theta_{13}$ and one Dirac type CP phase $\delta_{\rm CP}$. $m_1, m_2$ and $m_3$ are the masses of three neutrino mass eigenstates $\nu_1, \nu_2$ and $\nu_3$ respectively and $G_F$ is the Fermi constant. To see the effect of torsional couplings, we have done some simple algebraic manipulations.
The term involving the torsional couplings in eq. (\ref{eq:TDSE_for_3nu.a}) can be expressed as, 
\begin{eqnarray}
\begin{pmatrix}\lambda_1 & 0 & 0 \\ 0 & \lambda_2 & 0 \\ 0 & 0 & \lambda_3\end{pmatrix}\tilde{n}=
\begin{pmatrix}0 & 0 & 0 \\ 0 & \Delta\lambda_{21} & 0 \\ 0 & 0 & \Delta\lambda_{31}\end{pmatrix}\tilde{n} + {\lambda_1 \tilde{n} \mathbb{I}},
\label{lambda}
\end{eqnarray}
with $\Delta \lambda_{(2,3)1} = \lambda_{(2,3)} - \lambda_1$. Additionally,  we write $ \Delta \lambda_{(2,3)1} \tilde{n} =\Delta \lambda_{(2,3)1} n_e \lambda^{\prime}$ with $\lambda^{\prime}=\sum\limits_{f} \lambda_f \frac{n_f}{n_e}$, and denote
 $\Delta\lambda_{(2,3)1} \lambda^{\prime}$ as $\lambda_{(2,3)1}^{\prime}$ in unit of $G_F$. It should be noted that  as $\lambda'_{(2,3)1}$ depends on the difference between the couplings $\lambda_{2(3)}$ and $\lambda_1$, hence, it can have both positive and negative values depending on the sign of  $\lambda_{2(3)}-\lambda_1$.  After plugging eq. \ref{lambda} in eq. \ref{eq:TDSE_for_3nu.a}, we get the effective mass squared differences in presence of torsional couplings as, 
\begin{eqnarray}
    \Delta\tilde{m}_{31}^2 = \Delta m_{31}^2 + 2 E \tilde{n} \Delta \lambda_{31} = \Delta m_{31}^2 + 2 E n_e \lambda_{31}^{\prime}\;,\nonumber \\
    \Delta\tilde{m}_{21}^2 = \Delta m_{21}^2 + 2 E \tilde{n} \Delta \lambda_{21} = \Delta m_{21}^2 + 2 E n_e \lambda_{21}^{\prime} \;,
\end{eqnarray}
where $\Delta m^2_{(3,2)1} = m_{(3,2)}^2 - m_1^2$. It is important to notice that, the extra terms in the mass squared differences  depend on energy. Hence,  $\Delta \tilde{m}_{21}^2$ and $\Delta \tilde{m}_{31}^2$ will vary with energy even for the fixed  values of $\lambda_{21}^{\prime}$ and $\lambda_{31}^{\prime}$. Here it is also interesting to note that in this scenario oscillations of neutrinos are possible even when neutrinos are massless \cite{DeSabbata:1981ek}. 

To derive the neutrino oscillation probabilities, we have extended the methodology given in Ref.~\cite{Akhmedov:2004ny} for torsion. Thus we obtain the following expressions of appearance and disappearance probabilities for neutrinos: 
 \begin{align}
     P_{\nu_{\mu} \rightarrow \nu_e} &= 4 \sin^2 \theta_{13} \sin^2 \theta_{23} \frac{\sin^2 (\tilde{A}-1) \tilde\Delta}{(\tilde{A} -1 )^2}  \nonumber \\ &+ 2 \tilde\alpha \sin \theta_{13} \sin 2 \theta_{12} \sin 2 \theta_{23} \cos (\tilde\Delta + \delta_{CP}) \frac{\sin \tilde{A} \tilde\Delta}{\tilde{A}} \frac{\sin (\tilde{A}-1) \tilde\Delta}{(\tilde{A} -1 )} \nonumber \\ &+ \tilde{\alpha}^2 \sin^2 2 \theta_{12} \cos^2 \theta_{23} \frac{\sin^2 \tilde{A} \tilde\Delta}{\tilde{A}^2} 
     \equiv P_1 + P_2 + P_3 \;, \label{app} \\
     P_{\nu_{\mu} \rightarrow \nu_{\mu}}&= 1-\sin^2 2 \theta_{23} \sin^2 \tilde\Delta + \rm higher~ order~ terms\;,
    \label{dis}
 \end{align}
where $\tilde\Delta = \frac{ \Delta \tilde m_{31}^2 L}{4 E}$, $\tilde\alpha = \frac{\Delta \tilde m_{21}^2}{\Delta \tilde m_{31}^2}$ and $\tilde{A} = \frac{2 \sqrt{2} G_F n_e E }{\Delta \tilde m_{31}^2}$.
From these equations, we can see that, when we consider non-zero value of $\lambda_{21}^{\prime}$ (keeping $\lambda_{31}^{\prime}$=0), $\tilde\alpha$ gets changed, while $\tilde\Delta$ and $\tilde{A}$ remain same as their standard-interaction values. On the other hand,  in the presence of $\lambda_{31}^{\prime}$ (keeping $\lambda_{21}^{\prime}$=0), all the three parameters i.e., $\tilde\alpha$, $\tilde\Delta$ and  $\tilde{A}$ get modified. One more interesting point to notice is that for non-zero value of $\lambda_{31}^{\prime}$, we obtain a term in $\tilde\Delta$ which is directly proportional to the baseline $L$ and independent of the neutrino energy $E$. Therefore, the impact of $\lambda_{31}^{\prime}$ will be more pronounced in P2SO than DUNE as it has a baseline nearly twice of DUNE.

For our convenience, we split the expression for the appearance channel in three terms: $P_1, P_2$ and $P_3$. Among these three terms, $P_1$ is the leading order term whereas, $P_2$ and $P_3$ are the first  and second order sub-leading terms in $\tilde\alpha$ respectively. It is interesting to note that, $P_1$ is independent of $\lambda_{21}^{\prime}$ (there is no $\tilde\alpha$ term in $P_1$) whereas $P_3$ is independent of $\lambda_{31}^{\prime}$. In $P_3$, both $\tilde\alpha$ and $\tilde{A}$ are present in such a way that  the effect of $\lambda_{31}^{\prime}$ gets canceled. Therefore, when $\lambda_{21}^{\prime}$ is non-zero (keeping $\lambda_{31}^{\prime}$=0), $P_1$ does not deviate from its value corresponding to the standard oscillation (SO) case, but the values of $P_2$ and $P_3$ change from their SO values. As a result, the value of total probability $P_{\mu e}$ gets changed from its SO value. Now when $\lambda_{31}^{\prime}$ is non-zero (keeping $\lambda_{21}^{\prime}$=0), $P_1$ and $P_2$ terms are modified but $P_3$ remains unchanged. However, in this case, the change in total probability due to $\lambda_{31}^{\prime}$ is small as compared to the case when  $\lambda_{21}^{\prime}$ is non-zero. Therefore, the parameter $\lambda_{21}^{\prime}$ is expected to have higher sensitivity in the appearance channel probability as compared to $\lambda_{31}^{\prime}$. To better understand the observations mentioned above, in table~\ref{values}, we have given the numerical value of each term in the appearance channel probability. These numbers are calculated for an energy value of 2.5 (5.1) GeV which corresponds to the oscillation maximum in DUNE (P2SO). The table includes four different scenarios: SO case, only non-zero $\lambda_{21}^{\prime}$ case, only non-zero $\lambda_{31}^{\prime}$ case, and both non-zero $\lambda_{21}^{\prime}$ and $\lambda_{31}^{\prime}$ case. The values of probabilities are calculated using the oscillation parameter values from Table \ref{osc}. For the appearance channel, this table displays the numerical values of $P_1, P_2$, $P_3$ and their combination.  From this table we can understand that the deviation of the probability from SO values due to $\lambda_{21}^{\prime}$ is more as compared to $\lambda_{31}^{\prime}$. 

The expression for disappearance channel contains leading order term followed by some higher order terms. In the leading order term, only $\tilde\Delta$ is present which is independent of $\lambda_{21}^{\prime}$. However, the higher order terms, which are functions of $\tilde\alpha$ and $\tilde\alpha^2$ can have the effect of $\lambda_{21}^{\prime}$. Therefore, when only $\lambda_{21}^{\prime}$ is non-zero, the main effect in $P_{\mu \mu}$ will come from the higher order terms, and when only $\lambda_{31}^{\prime}$ is non-zero, the main contribution will come from the leading order term. This is obvious that the contribution from leading order term in presence of $\lambda_{31}^{\prime}$ will be more than the contribution from higher order terms in presence of $\lambda_{21}^{\prime}$. From Table \ref{values} which also lists the value of the  disappearance channel probability in its leading order, we see that the deviation of the probability from the SO value occurs only due to $\lambda_{31}^{\prime}$.

In the following sections, we will use the observations to explain our numerical results. 

\begin{table}[]
    \centering
    \begin{tabular}{||c||c||c||c||c||c||}
    \hline
    \hline
       Condition & $P_1$  & $P_2$   &  $P_3$   &  $P_1+P_2+P_3$  & Disappearance (leading order)   \\
          &   DUNE  & DUNE   &  DUNE  & DUNE  &  DUNE  \\
          &  (P2SO) & (P2SO)  & (P2SO)  &  (P2SO) & (P2SO) \\
        \hline
        \hline
        $\lambda_{21}^{\prime}=\lambda_{31}^{\prime}=0$ 
 &   0.0590 & 0.0130  &  0.0010 &  0.0731 & 0.0180  \\
   (SO) &  (0.0770)  &  (0.0133)  &  (0.0009) &  (0.0913)  &  (0.0133) \\
    \hline
    \hline
    $\lambda_{21}^{\prime}=0.1 ~G_F, \lambda_{31}^{\prime}=0$  & 0.0590 & 0.0194  & 0.0023  & 0.0808  &  0.0180  \\
       &  (0.0770)  &  (0.0267)  & (0.0036)  & (0.1074)   &  (0.0133)\\
\hline
\hline
$\lambda_{21}^{\prime}=0.0, \lambda_{31}^{\prime}=0.1 ~G_F$   & 0.0593 & 0.0132  & 0.0010 & 0.0736  &   0.0225 \\
& (0.0789)  & (0.0140)  & (0.0009)  &  (0.0939)  &  (0.0202)\\
\hline
\hline
$\lambda_{21}^{\prime}=0.1 ~G_F, \lambda_{31}^{\prime}=0.1 ~G_F$  &  0.0593 & 0.0197  & 0.0023  &  0.0814  
   &  0.0225 \\
  &  (0.0789)  &  (0.0280)  &  (0.0036) & (0.1106) &  (0.0202) \\
  \hline
  \hline
    \end{tabular}
    \caption{Numerical values of appearance and disappearance probabilities of DUNE (P2SO) experiment for four different conditions: SO, non-zero $\lambda_{21}^{\prime}$, non-zero $\lambda_{31}^{\prime}$ and both non-zero $\lambda_{21}^{\prime}$ and $\lambda_{31}^{\prime}$.}
    \label{values}
\end{table}

\section{Experimental setup and simulation}
\label{experi}

In this article, we have taken two future long-baseline neutrino experiments DUNE and P2SO. DUNE is one of the most promising upcoming long-baseline neutrino experiments with a baseline 1300 km from Fermi National Accelerator Laboratory (FNAL) to Sanford Underground Research Facility (SURF). For simulation of DUNE experiment, we have used the official files corresponding to the technical design report \cite{DUNE:2021cuw}. The experiment will have four liquid argon time-projection chamber (LArTPC) detector each of volume 10 kt with a beam power of 1.2 MW. In our simulation, we have taken 13 years of total run-time comprising of 6.5 years in neutrino mode and 6.5 years in anti-neutrino mode. 

The Protvino to Super-ORCA (P2SO) is also another future long-baseline neutrino experiment with baseline 2595 km. For simulation in P2SO experiment, we have used the technical design report of Refs.~\cite{Akindinov:2019flp, Hofestadt:2019whx}. Additionally, we refer to Refs.~\cite{Singha:2022btw,Majhi:2022fed,Singha:2023set} for the detailed configuration of the P2SO experiment. It will have a few Mt of fiducial detector volume with a beam power of 450 KW corresponding to $4 \times 10^{20}$ proton-on-target (POT). For simulation purpose, we have taken 3 years of neutrino mode and 3 years of anti-neutrino mode, a total 6 years of run-time.

To simulate DUNE and P2SO, we use General Long-Baseline Experiment Simulator (GLoBES) \cite{Huber:2004ka, Huber:2007ji} software. We have modified the probability engine to include the effect of torsion. This engine calculates the exact neutrino oscillation probabilities in matter in presence of torsion. For the determination of sensitivities, we use Poisson log-likelihood formula, 
\begin{equation}
    \chi^2 = 2 \sum_{i=1}^n \left[N_i^{\rm test} - N_i^{\rm true} - N_i^{\rm true} \rm{log} \left( \frac{N_i^{ \rm test}}{N_i^{\rm true}} \right) \right],
    \label{chi}
\end{equation}
where, $N_{i}^{\rm true}$ and $N_{i}^{\rm test}$ are the event numbers in true and test spectra respectively, `$i$' is the number of energy bins. The true values of the oscillation parameters are given in table \ref{osc} and are taken from Ref.~\cite{Esteban:2020cvm}. The relevant oscillation parameters are minimized in the $\chi^2$ analysis. we present our results only for the normal ordering of the neutrino masses i.e., $\Delta m^2_{31} > 0$.

\begin{table}[]
    \centering
    \begin{tabular}{||c||c||}
    \hline
    \hline
       Oscillation parameters  & True values \\
       \hline
       \hline
        $\theta_{12}$ & $33.41^{\circ}$\\
        \hline
        \hline
        $\theta_{13}$ & $8.58^{\circ}$ \\
        \hline
        \hline
        $\theta_{23}$  &  $42^{\circ}$ \\
        \hline
        \hline
        $\delta_{CP}$  &  $230^{\circ}$ \\
        \hline
        \hline
        $\Delta m_{21}^2  ~(eV^2)$  & $7.41 \times 10^{-5}$ \\
        \hline
        \hline
        $\Delta m_{31}^2 ~(eV^2)$  & $\pm 2.507 \times 10^{-3}$\\
        \hline
        \hline
    \end{tabular}
    \caption{Oscillation parameters and their values. We have used these values of their corresponding oscillation parameters as true values for our calculation.}
    \label{osc}
\end{table}

\section{Effect of torsional couplings on probability}
\label{tor-prob}

\begin{figure}[t!]
     \includegraphics[height=55mm, width=80mm]{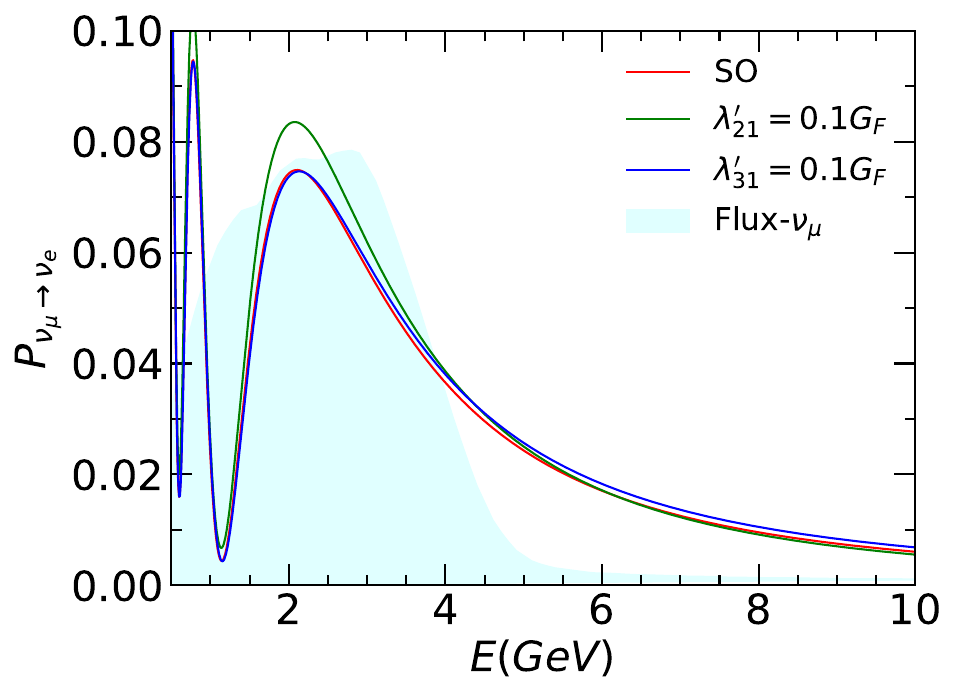}
     \includegraphics[height=55mm, width=80mm]{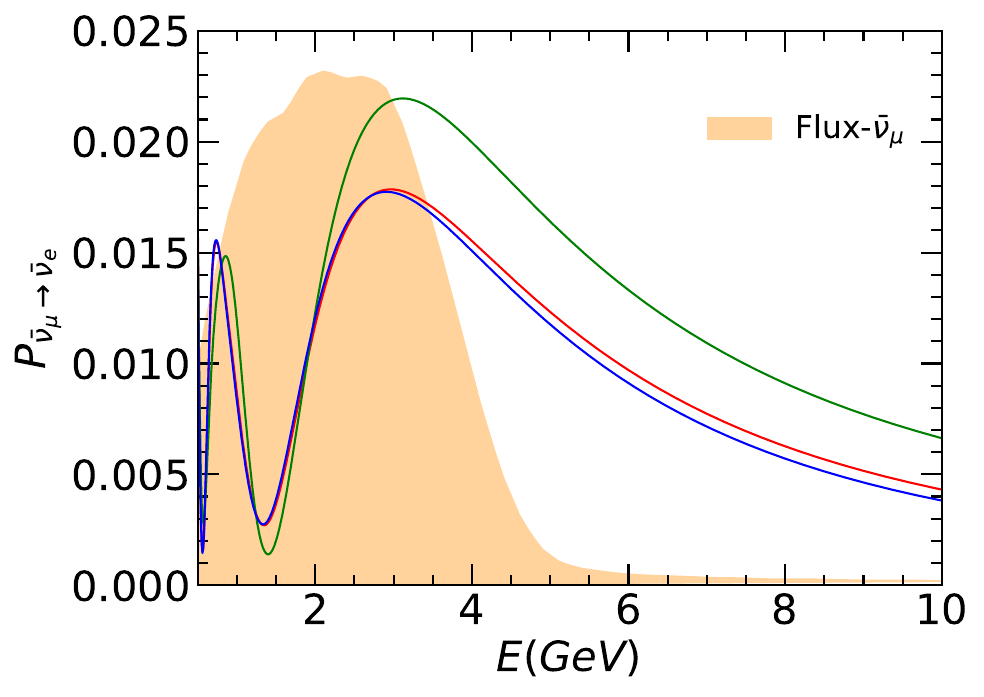}\\
     \includegraphics[height=55mm, width=80mm]{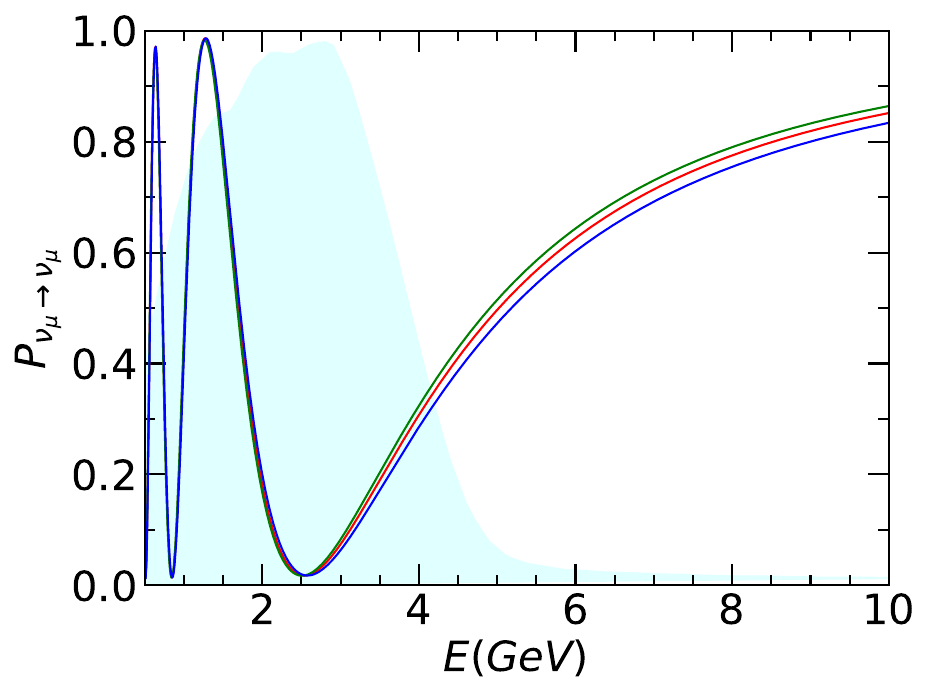}
     \includegraphics[height=55mm, width=80mm]{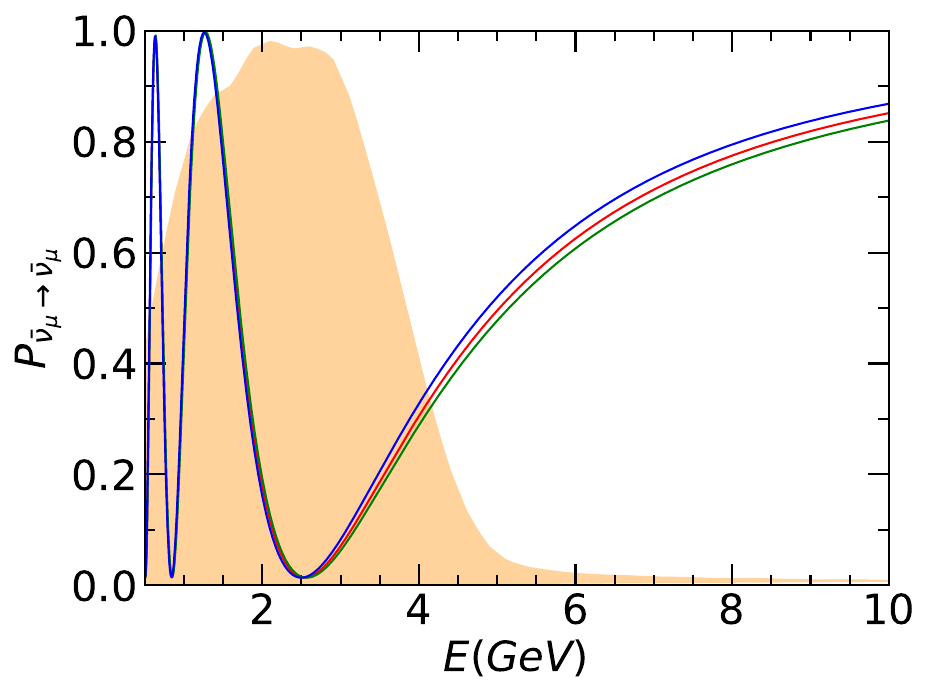}
     \caption{Appearance and disappearance probability for DUNE. Upper [lower] row shows the appearance [disappearance] probability in neutrino (left panel) and antineutrino (right panel) mode. In each panel, red, green and blue curves are for standard oscillation (SO) case, with taking $\lambda_{21}^{\prime}=0.1~G_F$ and $\lambda_{31}^{\prime}=0.1~G_F$ respectively. Cyan (brown) shaded region shows the corresponding fluxes of the respective experiments.}
     \label{prob-dune}
 \end{figure}

In this section, we will see the effect of torsional couplings $\lambda_{21}^{\prime}$ and $\lambda_{31}^{\prime}$ on the appearance and disappearance probabilities for the DUNE and P2SO baselines.  

 \begin{figure}
      \includegraphics[height=55mm, width=80mm]{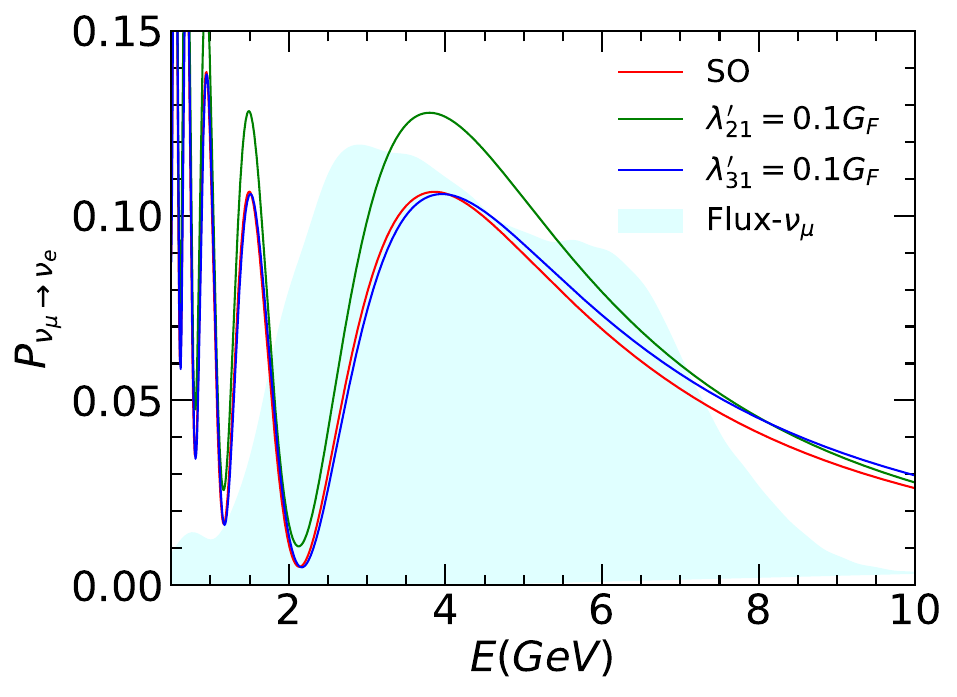}
      \includegraphics[height=55mm, width=80mm]{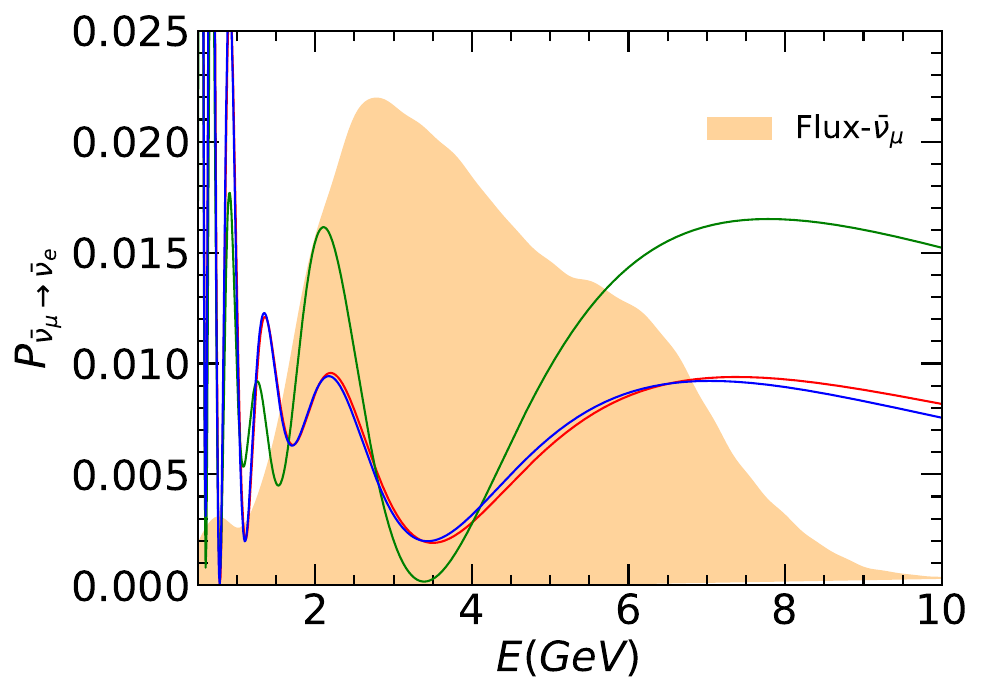}\\
      \includegraphics[height=55mm, width=80mm]{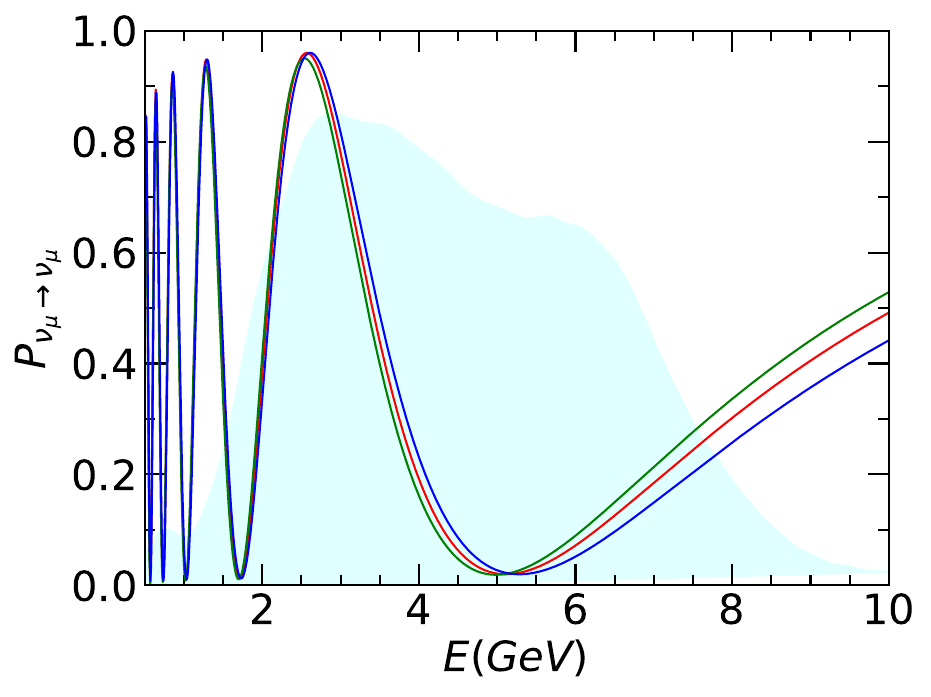}
      \includegraphics[height=55mm, width=80mm]{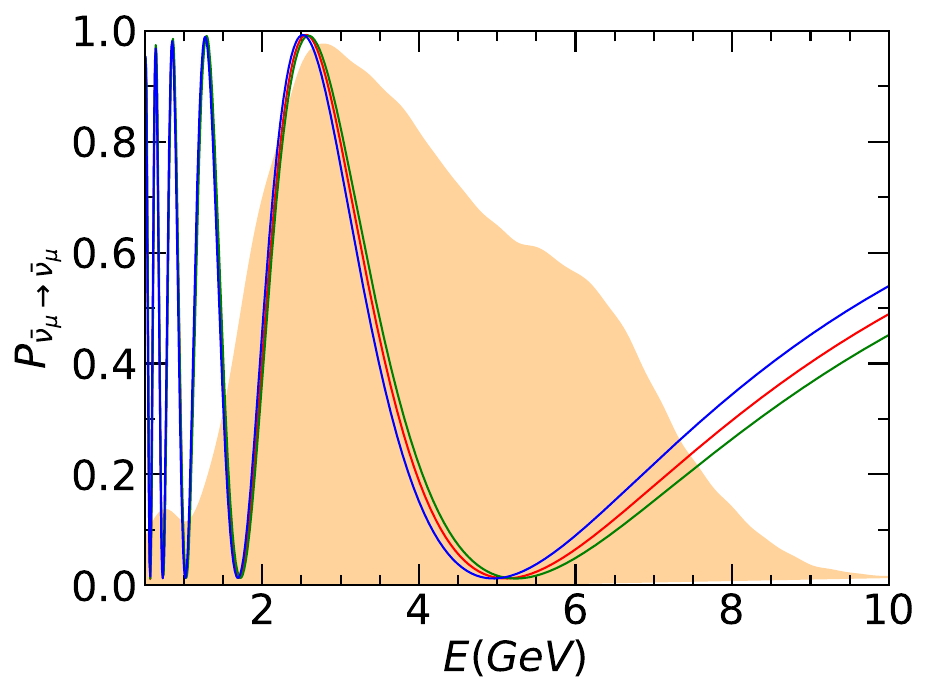}
      \caption{Same as \ref{prob-dune} but for P2SO experiment.}
      \label{prob-nue}
  \end{figure}
  
Let us first discuss the effect on appearance probability. To see the effect of the new couplings, we have plotted appearance probability with respect to neutrino energy in the upper panel of Figs. \ref{prob-dune} and \ref{prob-nue}. Fig. \ref{prob-dune} is for DUNE and Fig. \ref{prob-nue} is for P2SO experiment respectively. The top left (right) panel of both the figures shows neutrino (antineutrino) appearance probability. In all panels, red, green and blue curves represent the probability plot for the standard oscillation case (SO), $\lambda_{21}^{\prime}=0.1~G_F$ and $\lambda_{31}^{\prime}=0.1~G_F$ conditions respectively. From the panels, we can clearly see the features discussed in section \ref{3nu}. Indeed, we notice that the change in probability due to $\lambda_{21}^{\prime}$ is much larger as compared to $\lambda_{31}^{\prime}$. This behaviour is same for neutrinos and antineutrinos for both the experiments. 

\begin{figure}
    \includegraphics[height=55mm, width=80mm]{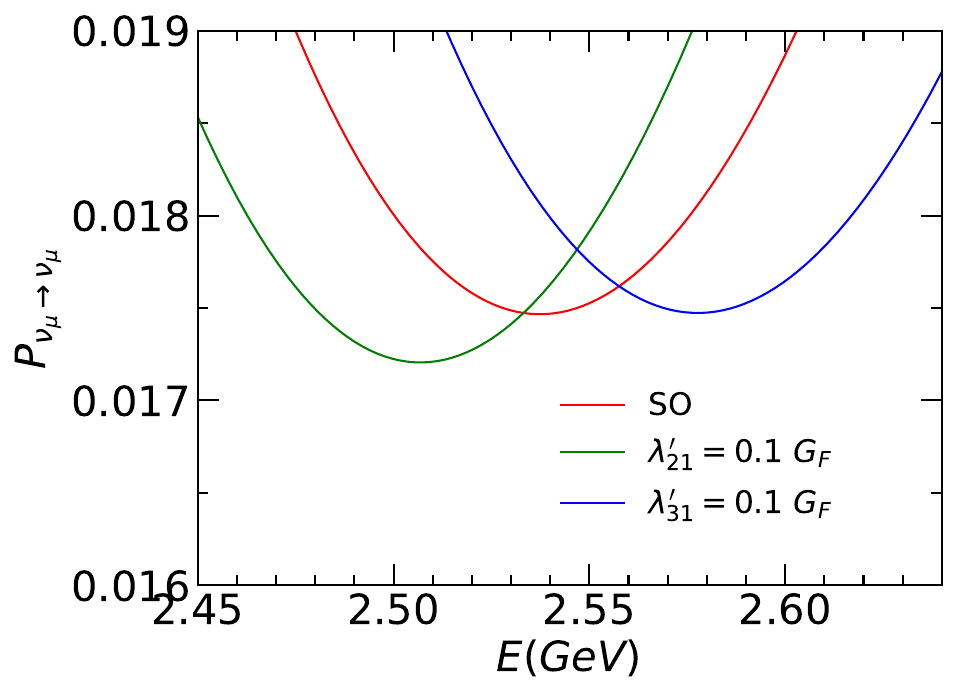}
    \includegraphics[height=55mm, width=80mm]{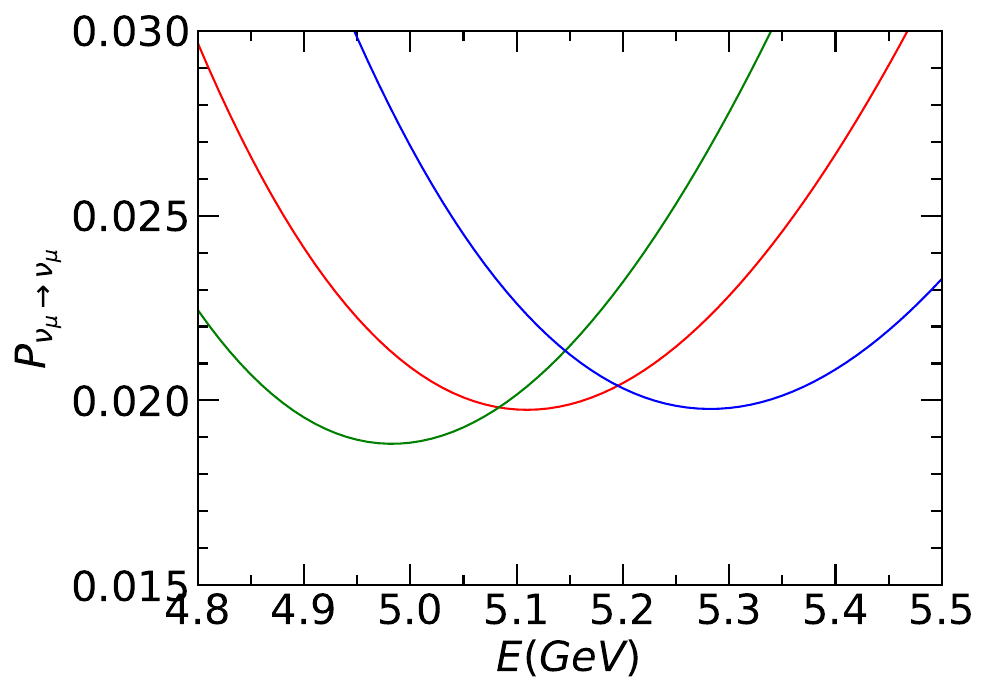}
    \caption{Disappearance probability at oscillation minima for DUNE (left) and P2SO (right).}
    \label{close}
\end{figure}

\begin{figure}
\includegraphics[height=55mm, width=80mm]{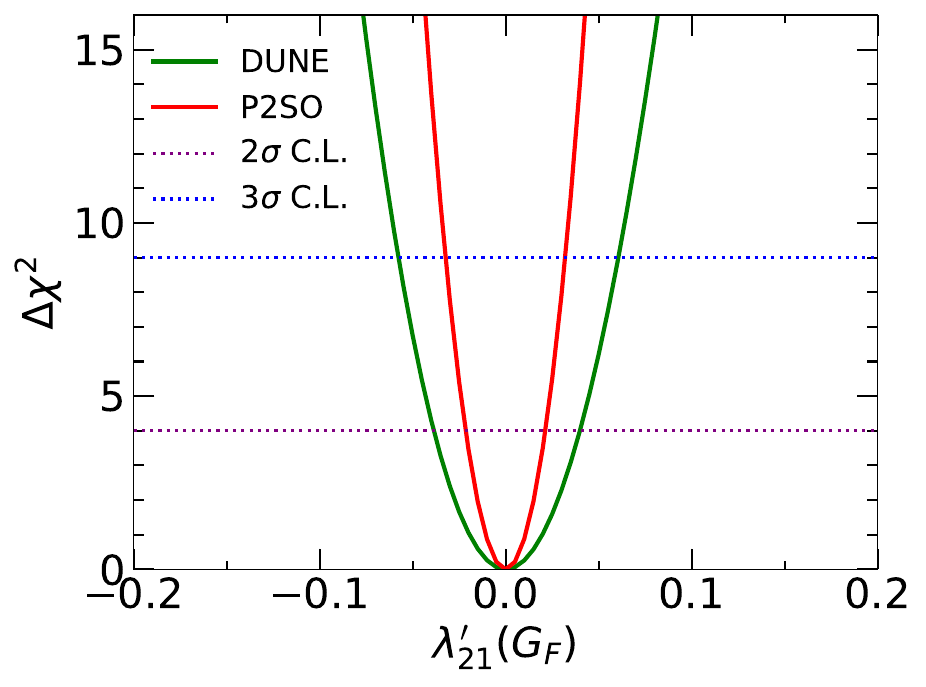}
     \includegraphics[height=55mm, width=80mm]{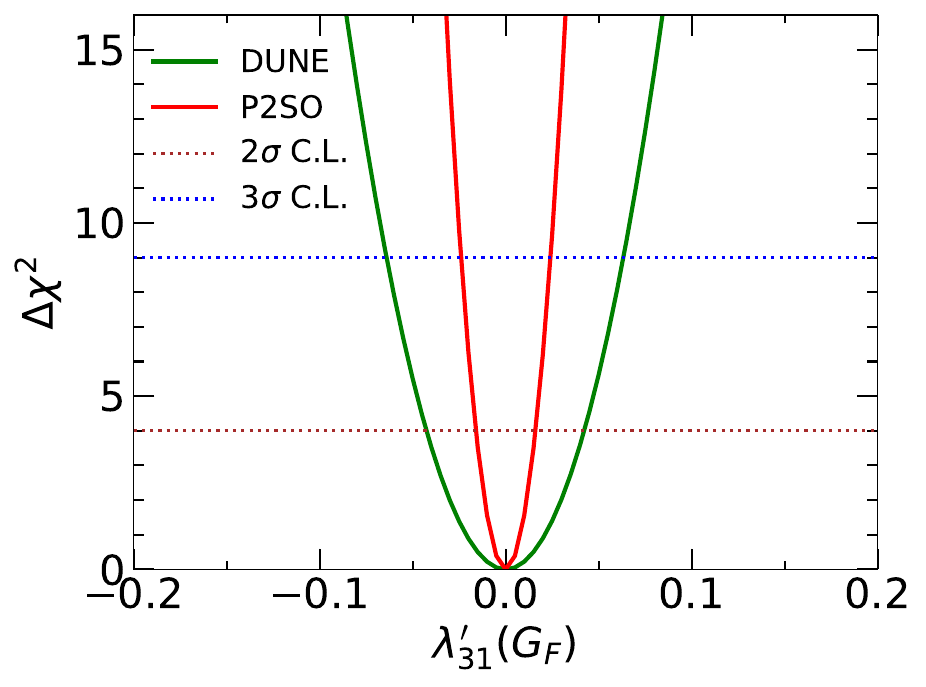}
     \caption{Left (right) panel shows the constraints of $\lambda_{21}^{\prime}~ (\lambda_{31}^{\prime})$ in the future long-baseline experiments. In each panel, red curve is for DUNE and green curve is for P2SO experiment. Purple (blue) dashed line represents the benchmark sensitivity of $2 \sigma~ (3 \sigma)$ C.L. }
     \label{bund}
 \end{figure}

Let's now discuss the disappearance probability. These are shown in the lower panels of Figs. \ref{prob-dune} and \ref{prob-nue}. In each figure, bottom left (right)  panel is for neutrino (antineutrino) mode. Color code for these panels are same as the upper row. Red, green and blue curves represent the disappearance probability for SO, $\lambda_{21}^{\prime}=0.1 ~G_F$ and $\lambda_{31}^{\prime}=0.1 ~G_F$ conditions respectively. From these panels, after taking a closer view, we can see that, the effect of $\lambda_{31}^{\prime}$ is more in disappearance probability than $\lambda_{21}^{\prime}$ confirming the observations of section \ref{3nu}. This conclusion is true for both DUNE and P2SO experiments. 
 
As the oscillatory terms in the probability expressions depends on torsion, it can change the position of the oscillation maximum/minimum. We have shown this in Fig. \ref{close} for the disappearance channel. In this figure, we have zoomed-in the location of the minima of the disappearance probabilities corresponding to neutrinos for DUNE (left panel) and P2SO (right panel) experiment. For both the panels, we can see a significant shift in the oscillation minima in presence of new torsional couplings. One can observe similar behaviour in the appearance channel too. However, these changes are quite minimal to be visible as seen from Figs. \ref{prob-dune} and \ref{prob-nue}.

\section{Bounds on torsional couplings}
\label{bound-tor}

In this section, we will study the capability of DUNE and P2SO to put bounds on the new torsional couplings. It should be noted that the experimental bounds on torsional couplings in electron-electron and lepton-quark sectors are presented in Ref.~\cite{Chakraborty:2024zek}. In our analysis, to obtain the bound on the torsional couplings in the neutrino sector, we take the standard three flavor oscillation scenario in the true spectrum of the $\chi^2$ and torsion in the test spectrum of the $\chi^2$. Here we have presented our results as a function of the test values of the torsional coupling constants.

 \begin{figure}
     \centering
     \includegraphics[height=65mm, width=80mm]{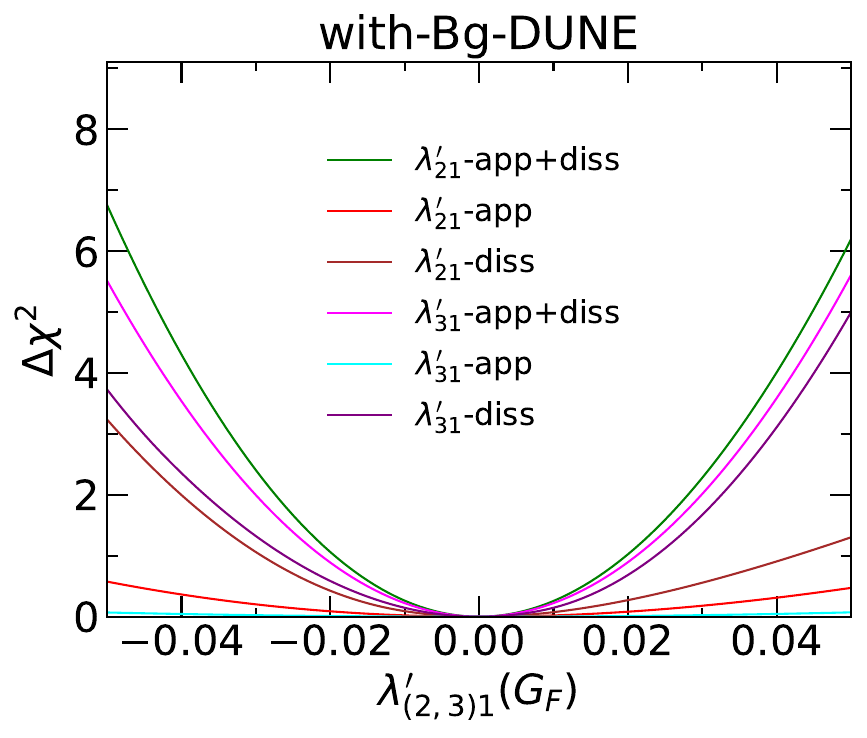}
     \includegraphics[height=65mm, width=80mm]{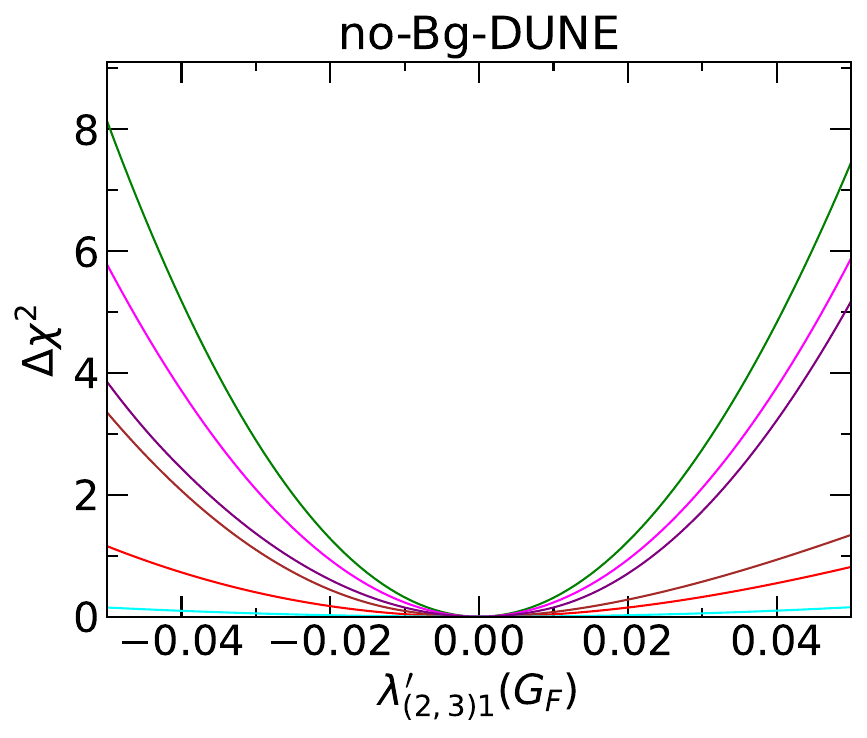}\\
     \includegraphics[height=65mm, width=80mm]{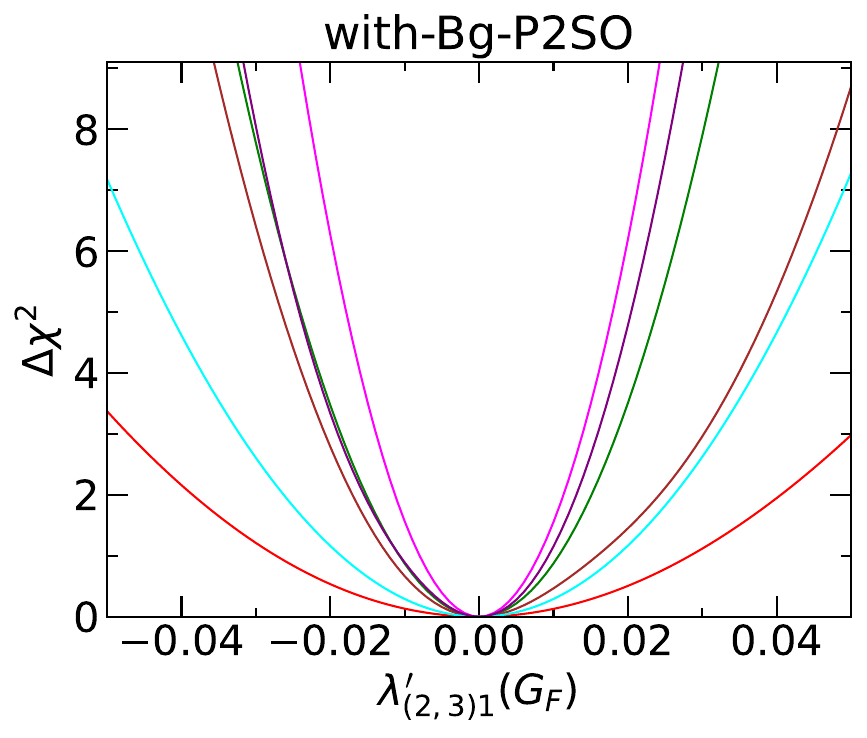}
     \includegraphics[height=65mm, width=80mm]{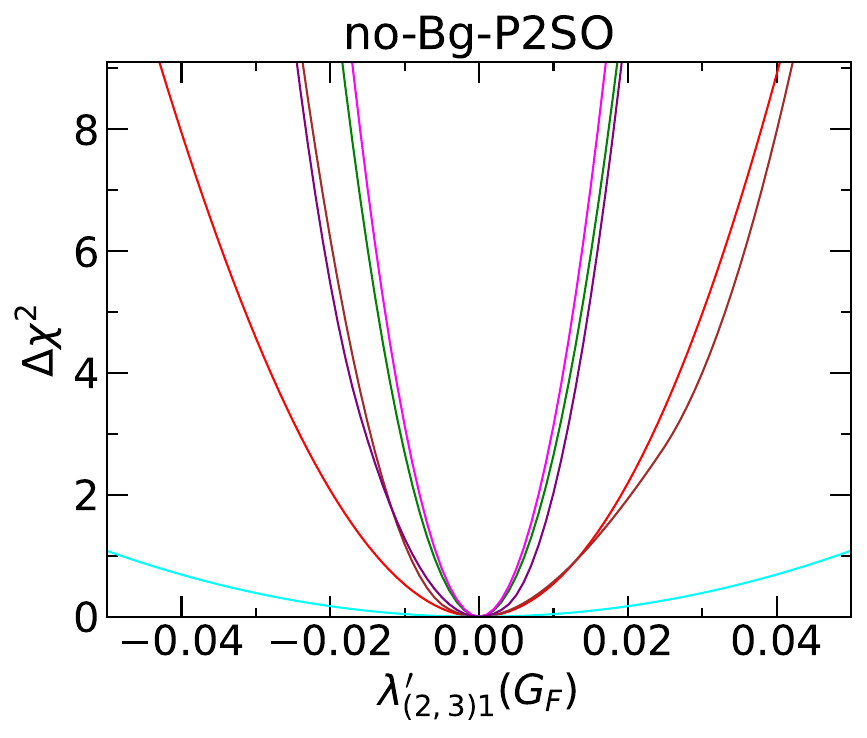}
     \caption{Sensitivity plots as a function of new torsional couplings. Upper (lower) row is for DUNE (P2SO) experiment. Left (right) column is the case when background is present (absent). Color codes are given in the legend. ``app" (``diss") [``app+diss"] stands for the condition when only appearance (disappearance) [appearance and disappearance both] channel is taken in consideration.}
     \label{bound-app}
 \end{figure}

 \begin{table}[]
    \centering
    \begin{tabular}{||c||c||c||c||c||}
    \hline
    \hline
       Experimental Setup  &  $\lambda_{21}^{\prime}$ bound  [$2 \sigma~ (3 \sigma) $]  & $\lambda_{31}^{\prime}$ bound [$2 \sigma~ (3 \sigma) $] & $\lambda_{21}^{\prime}$ bound  & $\lambda_{31}^{\prime}$ bound \\
  & (when $\lambda_{31}^{\prime}$ = 0) & (when $\lambda_{21}^{\prime}$ = 0) & (when $\lambda_{31}^{\prime} \neq 0$)  & (when $\lambda_{21}^{\prime} \neq 0$) \\
          \hline
          DUNE & $-0.039 ~(-0.057) $  &  $-0.043 ~(-0.064) $& $ -0.060 $  & $ -0.070 $ \\
          \hline 
             &   0.040 (0.060)     & 0.043 (0.063) & 0.080  &  0.080   \\
             \hline
             \hline
             P2SO & $ -0.022 ~
 (-0.032)$  & $-0.016 ~(-0.024)$  &  $-0.080 $  & $ -0.060 $  \\
             \hline
               &     0.022 (0.032)  &  0.016 (0.024) & 0.130  &  0.100 \\
               \hline
                     \hline
    \end{tabular}
    \caption{$\lambda_{21}^{\prime}$ and $\lambda_{31}^{\prime}$ bounds (in the units of $G_F$) for different experiments, DUNE and P2SO. Second (third) column of the table shows $\lambda_{21}^{\prime} (\lambda_{31}^{\prime})$ bound for these two experiments at $2 \sigma ~(3 \sigma)$ C.L. by taking one parameter at a time. Fourth (fifth) column gives the bound on $\lambda_{21}^{\prime} (\lambda_{31}^{\prime})$ at $2 \sigma$ C.L. by taking both couplings at the same time. }
    \label{l21-bound}
\end{table}

First let us consider one parameter at a time. Fig.~\ref{bund} shows the bounds of $\lambda_{21}^{\prime}$ and $\lambda_{31}^{\prime}$ in DUNE and P2SO experiments and the results are summarized in table \ref{l21-bound}. Left panel is for $\lambda_{21}^{\prime}$ and right panel is for $\lambda_{31}^{\prime}$. In each panel,  green and red curves are for DUNE and P2SO experiments respectively. Purple (blue) dotted curve is the benchmark value of $2 \sigma~ (3 \sigma)$ C.L. From these panels, we can see that, P2SO gives more stringent bound on both the parameters than DUNE. From our earlier discussion, we have understood that sensitivity of $\lambda_{21}^{\prime}$ comes from the appearance channel whereas sensitivity for $\lambda_{31}^{\prime}$ comes from the disappearance channel. Therefore, a combination of appearance and disappearance channel will be sensitive to both $\lambda_{21}^{\prime}$ and $\lambda_{31}^{\prime}$. Between these two parameters, interestingly, DUNE gives better bound on $\lambda_{21}^{\prime}$ while P2SO gives better bound on $\lambda_{31}^{\prime}$. To understand this discrepancy, in Fig. \ref{bound-app} we tried to look at the effect of backgrounds in P2SO and DUNE. Here it is important to note that the Super-ORCA detector in P2SO is designed to study neutrinos for astrophysical source and therefore this detector is not optimized for the beam based backgrounds. In Ref.~\cite{Singha:2021jkn} it was shown that the large background (mostly misidentified muons) of the ORCA detector can significantly impact the physics sensitivity of the Protvino to ORCA experiment.

To explain the behavior of Fig. \ref{bund}, we have plotted Fig. \ref{bound-app}, where we have shown the effect of background on appearance, disappearance and the combined channels explicitly. In this figure, upper (lower) row is for DUNE (P2SO) experiment. Left (right) column shows the sensitivity plot of new torsional couplings with (without) backgrounds. In each panel, green, red and brown curves are for the sensitivity of $\lambda_{21}^{\prime}$ and magenta, cyan and purple are for the sensitivity of $\lambda_{31}^{\prime}$. For each panel, ``app" (``diss") represents the sensitivity results by taking only appearance (disappearance) channel and ``app+diss" stands for the sensitivity by taking both appearance and disappearance channel together. From the panels we see the following. Let us first consider the appearance channel. When there is no background, we see that the sensitivity of $\lambda_{21}^{\prime}$ (red curve) is higher than the sensitivity of $\lambda_{31}^{\prime}$ (cyan curve) for both DUNE and P2SO. However, when the background is turned on, the sensitivity of $\lambda_{21}^{\prime}$ becomes less than $\lambda_{31}^{\prime}$ in P2SO which is not the case in DUNE. This signifies the fact that though appearance channel has a better sensitivity to $\lambda_{21}^{\prime}$, the presence of background, significantly limits its capacity for P2SO experiment. 
 
Now let us see the disappearance channel. In this case, the sensitivity of $\lambda_{31}^{\prime}$ (purple curve) is always greater than the sensitivity of $\lambda_{21}^{\prime}$ (brown curve) in both DUNE and P2SO irrespective of the effect of background. 
Now, when both appearance and disappearance channels are taken together, the sensitivity of $\lambda_{31}^{\prime}$ (magenta curve) becomes greater than the sensitivity of $\lambda_{21}^{\prime}$ (green curve) for P2SO experiment when the backgrounds are turned on. Whereas for DUNE, the combination of the appearance and disappearance channel is always dominated by $\lambda_{21}^{\prime}$ irrespective of the background.

\begin{figure}
     \includegraphics[height=70mm, width=90mm]{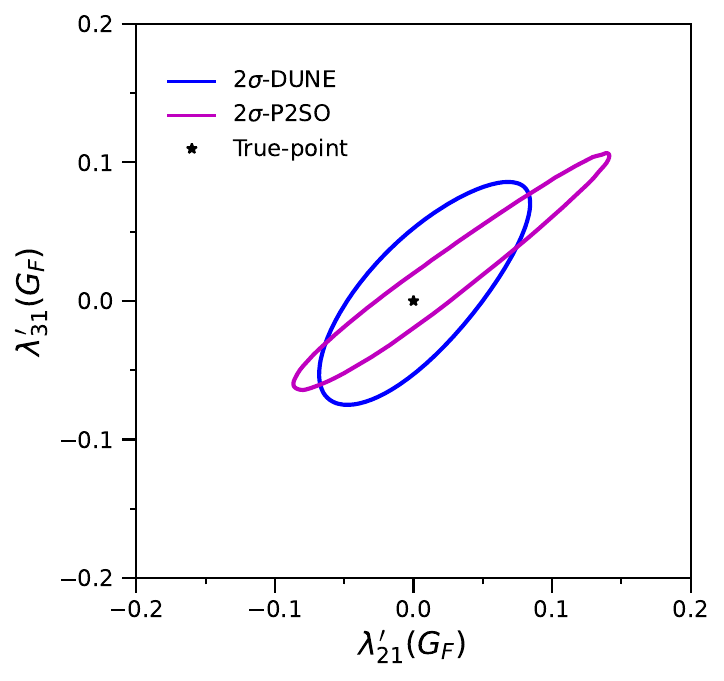}
     \caption{Two dimensional bound plot of $\lambda_{21}^{\prime}$ and $\lambda_{31}^{\prime}$ for two experiments: blue is for DUNE and magenta is for P2SO experiment at $2 \sigma$ C.L. Black star indicates the Standard oscillation (SO) case. }
     \label{dune-t2hk-t2hkk}
 \end{figure}

Now let us consider the case when both the torsion coupling constants are taken at the same time. In Fig. \ref{dune-t2hk-t2hkk}, we have shown contours in the  $\lambda_{21}^{\prime}-\lambda_{31}^{\prime}$ plane at $2 \sigma$ C.L. for two degrees of freedom in P2SO and DUNE experiments. In this figure, blue and magenta curves are for DUNE and P2SO respectively. The elliptic shape of the contours show a strong correlation between the two torsional coupling constants. Further we see that when one takes both the parameters at the same time, the upper bounds on these parameter weakens significantly compared to the case when one parameter is taken at a time. 

We have included the bounds at $2 \sigma$ C.L. obtained on the torsional coupling constants when both of them are taken at the same time in the fourth and fifth columns of table \ref{l21-bound}.
From this table, we see that, when we take both couplings at the same time, DUNE provides  similar bound on both couplings, but P2SO provides more constraint on $\lambda_{31}^{\prime}$ than on $\lambda_{21}^{\prime}$. When comparing the two experiments, DUNE gives better constraint on both the couplings than P2SO. Note that, the experimental bounds on torsional couplings in the electron-electron  and lepton-quark sectors are $|\lambda_e^V \lambda_e^A| \verb|<| 1.917 \times 10^{-8}$ GeV$^{-2}$ and $|\lambda^e \lambda^q| \verb|<| 1.429 \times 10^{-7}$ GeV$^{-2}$ respectively, which are of the same order of magnitude as our bounds from P2SO and DUNE in the neutrino sector.

In addition to the torsional coupling discussed above, there is another type, known as ``axial torsional coupling", is explored in Ref.~\cite{Obukhov:2014fta}. In their study, the authors have obtained a bound on the axial torsional coupling using data from experiments involving a Dirac particle in a uniform magnetic field and the Earth's rotating frame. Their analysis shows that torsion couples directly with the axial current, and this interaction appears in the Hamiltonian of a atomic system affecting the Zeeman frequencies for transitions between adjacent atomic levels. However, for neutrinos, to have the chiral properties (left-handed and right-handed states) we need the contributions from both axial and vector currents in the torsion field (as shown in the Appendix). It is well-established, particularly from studies of neutrino interactions in matter (MSW effects \cite{Wolfenstein:1977ue,Mikheev:1987qk}), that the axial current contribution is several orders of magnitude smaller than that of the vector current. For this reason, the bound obtained in Ref.~\cite{Obukhov:2014fta} is several order of magnitude stronger than ours. However, these two theories are completely different and therefore the bounds obtained on the torsional coupling constants from these two theories can't be compared.

\section{Physics sensitivities in presence of torsion}
\label{phys}

In this section, we will present the physics sensitivities of DUNE and P2SO in presence of torsion i.e., the effect of torsion in the determination of CP violation, mass ordering of the neutrinos and octant of $\theta_{23}$. We will also comment on our findings regarding the  precision measurement of the atmospheric parameters i.e., $\theta_{23}$ and $\Delta m^2_{31}$ in presence of torsion. In our analysis, we have estimated these sensitivities by taking the torsion scenario in both true and test spectrum of the $\chi^2$. In these calculations, we have kept the torsional couplings fixed in the test and plotted the sensitivity as a function of their true values. 

\begin{figure}[htbp]
     \includegraphics[height=70mm, width=81mm]{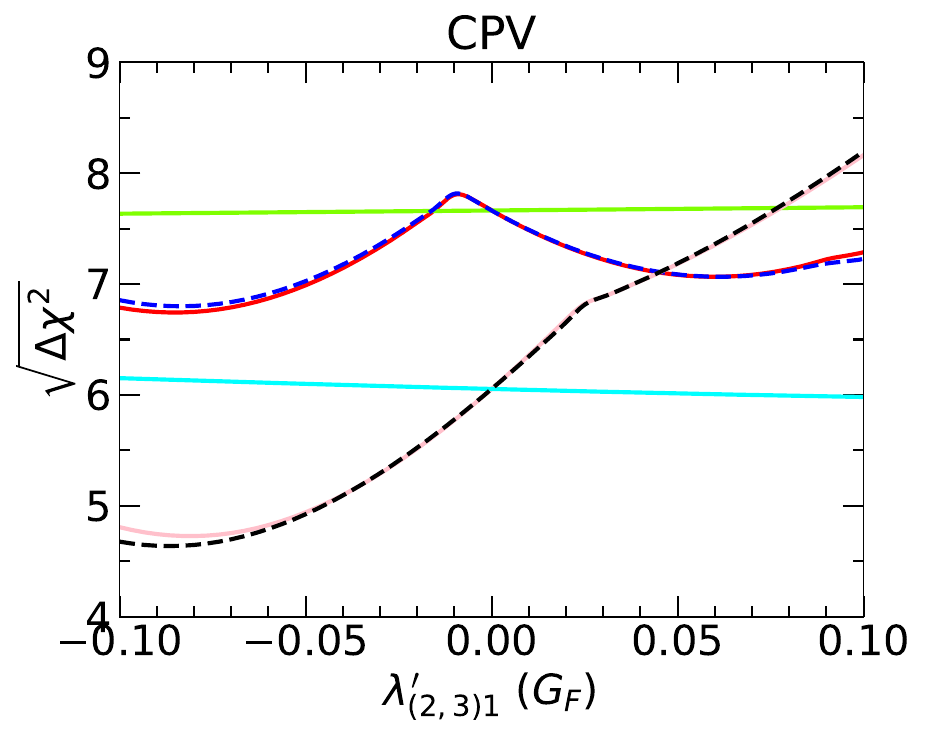}
     \includegraphics[height=70mm, width=81mm]{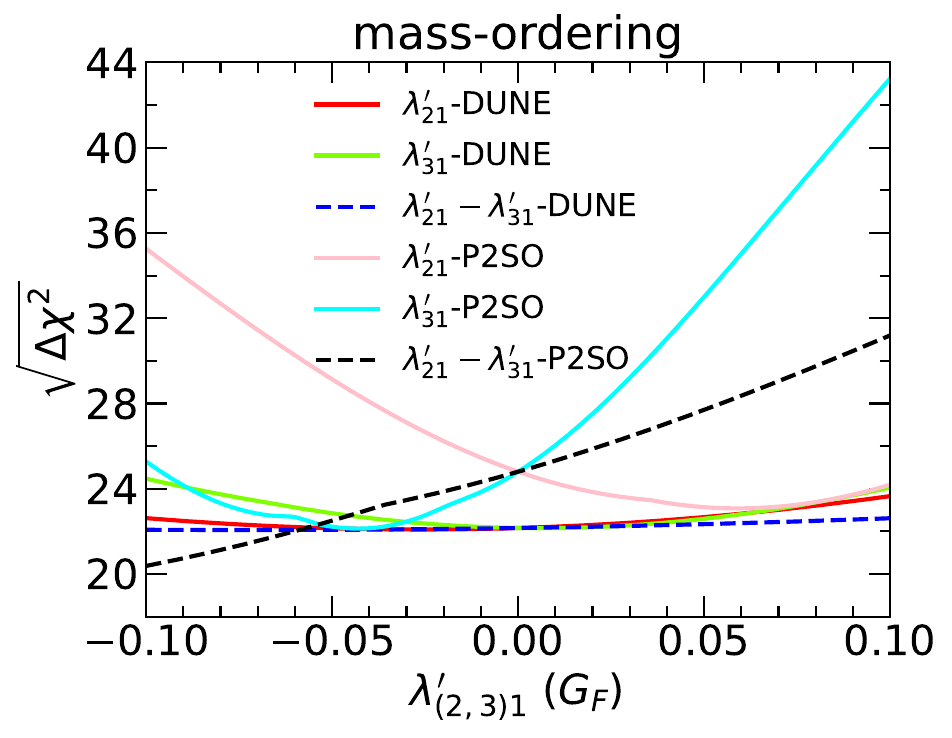}
     \includegraphics[height=70mm, width=81mm]{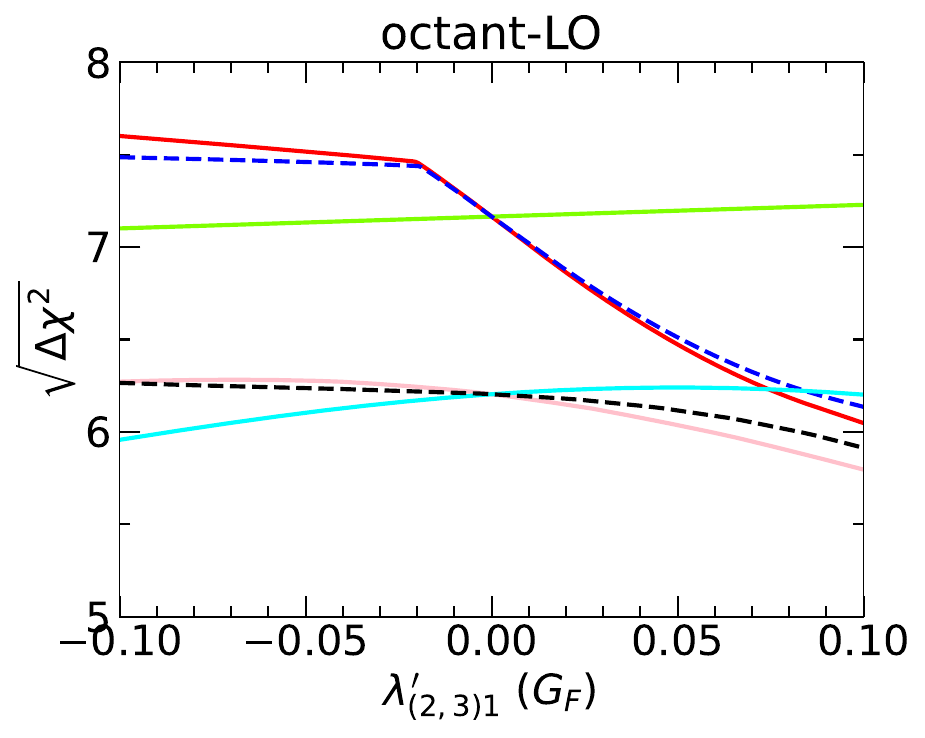}
     \caption{Physics sensitivities in presence of torsional couplings. Left (right) panel of upper row depicts the CPV (mass-ordering) sensitivity as a function of $\lambda_{(2,3)1}^{\prime}$ in units of $G_F$. Lower row shows the octant sensitivity of atmospheric mixing angle $\theta_{23}$ in variation with  $\lambda_{(2,3)1}^{\prime}$. Color codes are given in the legend.}
     \label{physics}
 \end{figure}
The presence of $\lambda_{21}^{\prime}$ and $\lambda_{31}^{\prime}$ can affect the CP violation (CPV) sensitivity results in P2SO and DUNE experiments. The deviation of the CPV sensitivity from standard oscillation scenario is shown in the top left panel of Fig. \ref{physics} for $\delta_{CP}^{\rm true} = 270^{\circ}$. We estimate the CPV sensitivity by taking two values of $\delta_{CP}$ ($0^{\circ}, 180^{\circ}$) in test spectrum and minimizing over them. In this panel, red solid, green solid and blue dashed curves represent the results for DUNE experiment whereas pink solid, cyan solid and black dashed curves show the results for P2SO experiment. Red and pink curves have been generated for $\lambda_{21}^{\prime}$, keeping $\lambda_{31}^{\prime}$ to zero. Similarly, green and cyan color curves are for $\lambda_{31}^{\prime}$, keeping $\lambda_{21}^{\prime}$ to zero. Blue and black dashed curves are the results when both the new parameters are taken at the same time. From the left panel of Fig. \ref{physics}, we can see the variation of CP violation sensitivity is more for $\lambda_{21}^{\prime}$, whereas the change is marginal for  $\lambda_{31}^{\prime}$ when we consider one parameter at a time. If we take both the parameters simultaneously, the result is very similar to the case when $\lambda_{21}^{\prime}$ is non-zero and $\lambda_{31}^{\prime}$ is zero. This is because, in the long-baseline experiments, the CP sensitivity mainly comes from the appearance channel which is more sensitive to $\lambda_{21}^{\prime}$ as compared to $\lambda_{31}^{\prime}$. Therefore, it is obvious that in presence of torsion, the CPV will be dominated by $\lambda_{21}^{\prime}$ when one considers both the parameters at the same time.  Even though the above mentioned features are applicable to both the experiments, there are some differences in the sensitivities in the results for these two experiment. In the left panel, for DUNE, we can see that, if we go far from zero value of $\lambda_{21}^{\prime}$, in both side i.e., negative as well as positive, CPV decreases. This is not true for P2SO. For P2SO, when we go from negative value of $\lambda_{21}^{\prime}$ to positive value, CPV increases all the time. We have checked that if we do not consider any background in P2SO, the behaviour of its CPV curve matches with DUNE.

Mass ordering sensitivity is the capability of any experiment to differentiate normal mass ordering from inverted ordering. In presence of new torsional coupling constants $\lambda_{21}^{\prime}$ and $\lambda_{31}^{\prime}$, the standard mass ordering sensitivity can change. In the top right panel of Fig. \ref{physics}, we have shown the mass ordering sensitivity for DUNE and P2SO. We have calculated this sensitivity by taking normal ordering in the true spectrum and inverted ordering in the test spectrum of the $\chi^2$. From the panel we see that, for DUNE, there is a marginal change in sensitivity in the presence of torsion. 

For P2SO, when we vary $\lambda_{21}^{\prime}$ from negative value to positive value keeping $\lambda_{31}^{\prime}$ to zero, the value of $\sqrt{\Delta \chi^2}$ decreases linearly. Exactly opposite nature is seen for the case of the variation of $\lambda_{31}^{\prime}$, keeping $\lambda_{21}^{\prime}$ to zero. 

When we vary both the new parameters simultaneously from negative value to positive one, the mass hierarchy increases linearly for both P2SO and DUNE.

Octant sensitivity is the capability of any experiment to resolve the degeneracy between lower octant (LO) and higher octant (HO). In presence of torsional couplings $\lambda_{21}^{\prime}$ and $\lambda_{31}^{\prime}$, octant sensitivity of P2SO and DUNE may change. In the bottom panel of Fig. \ref{physics}, we have shown the effect of $\lambda_{21}^{\prime}$ and $\lambda_{31}^{\prime}$ on the determination of octant sensitivity. For generating this panel, we have taken LO in the true spectrum and HO in test spectrum. For octant sensitivity we see that the change in sensitivity due to $\lambda_{31}^{\prime}$ is marginal. For $\lambda_{21}^{\prime}$ when we go from negative value of $\lambda_{21}^{\prime}$ to positive value, the sensitivity decreases. 

For the condition when both the couplings are varying simultaneously, the octant sensitivity decreases if we go from negative to positive value of coupling constants. Above mentioned results are true for both the experiments DUNE and P2SO.  

Finally, regarding the precision measurement of the atmospheric parameters i.e., $\theta_{23}$ and $\Delta m_{31}^2$, we have checked that the effect is almost negligible. So, even if we add torsion in the theory of neutrino oscillation, the precision of $\theta_{23}$ and $\Delta m_{31}^2$ remains unaltered.

\section{Summary and Conclusion}
\label{summary}

In this paper, we have studied the effect of gravity on neutrino oscillation in the context of two upcoming long-baseline experiments DUNE and P2SO. In the presence of curved spacetime, the Dirac equation gets modified. This modification gives rises to two torsional coupling constants i.e., $\lambda_{21}^{\prime}$ and $\lambda_{31}^{\prime}$, which affect the Hamiltonian of neutrino oscillation. 

In the first part of our paper, we have derived the neutrino oscillation probabilities in the presence of torsion. We have shown that the appearance channel is mainly sensitive on $\lambda_{21}^{\prime}$ whereas the sensitivity of $\lambda_{31}^{\prime}$ mainly comes from the disappearance channel. Our analysis shows that the presence of torsion can produce a small shift in the oscillation maximum/minimum. From the probability expressions, we also concluded that the impact of $\lambda_{31}^{\prime}$ will be more pronounced in P2SO as compared to DUNE. 

Next, we studied the capability of DUNE and P2SO to put bound on the torsional couplings. When we take one parameter at a time, we find that P2SO gives more stringent bound on both the parameters than DUNE. 
For DUNE experiment, among the two parameters, $\lambda_{21}^{\prime}$ is more constraint than $\lambda_{31}^{\prime}$ whereas for P2SO, $\lambda_{31}^{\prime}$ has better bound than $\lambda_{21}^{\prime}$.
This is because of the large background of P2SO which limits its capability to measure $\lambda_{21}^{\prime}$. However, if we take both the parameters at the same time, the upper bounds on these parameter weakens significantly. In this case we see that DUNE provides  similar bounds on both the couplings and its sensitivity is better than P2SO. 

Finally, we have shown the effect of torsion in the measurement of CP violation, mass ordering and octant. Our numerical results show that  variation of CP violation sensitivity is more for $\lambda_{21}^{\prime}$, whereas the change is marginal for $\lambda_{31}^{\prime}$. Therefore, when we consider both the parameters at the same time, the sensitivity is dominated by $\lambda_{21}^{\prime}$. We also noticed certain difference in the behaviours of the DUNE and P2SO curves, which are due to the large background of P2SO. For mass ordering sensitivity, the change in the sensitivity with respect to  both the torsional coupling constants are marginal in DUNE, whereas the change in the sensitivity is significant for P2SO. Regarding octant, the sensitivity of DUNE and P2SO are affected by $\lambda_{21}^{\prime}$ whereas change in the sensitivity due to $\lambda_{31}^{\prime}$ is marginal. In addition, we have also checked that the torsional coupling constants do not alter the precision of $\theta_{23}$ and $\Delta m_{31}^2$.

In summary, our results show that if we consider the effect of curved spacetime on neutrino oscillation, it can significantly affect the sensitivity of the neutrino oscillation experiments. 
Thus, it is very important to have a clear understanding of the concept of torsion and its impact on neutrino oscillation.

\section{Acknowledgements}
The work of MG has been in part funded by Ministry of Science and Education of Republic of Croatia grant No. KK.01.1.1.01.0001 and European Union under the NextGenerationEU Programme. Views and opinions expressed are however those of the author(s) only and do not necessarily reflect those of the European Union. Neither the European Union nor the granting authority can be held responsible for them. MG would like to thank Teppei Katori and Osamu Yasuda for useful discussions. PP and DKS want to thank Prime Minister’s Research Fellows (PMRF) scheme for its financial support. RM would like to acknowledge University of Hyderabad IoE project grant no. RC1-20-012. We gratefully acknowledge the use of CMSD HPC facility of University of Hyderabad to carry out the computational works.

\appendix

\section{Theoretical background} 
\label{append}

 We use  Einstein-Cartan-Sciama-Kibble (ECSK) formalism  of gravity \cite{Izaurieta:2020kuy,Esposito:1990ub,Szczyrba:1984fcb} for our study. This framework represents a first-order formulation of gravity, employing tetrad fields. The internal flat space metric and the spacetime metric are connected by tetrad fields ($e^{a}_{\mu}$). These expressions are as follows 
\begin{eqnarray}
\eta_{ab} e_{\mu}^a e_{\nu}^b = g_{\mu \nu}, ~~~~ g_{\mu \nu} e_{a}^{\mu} e_{b}^{\nu} = \eta_{ab}, ~~~~ e_{a}^{\mu} e_{\nu}^{a} = \delta_{\nu}^{\mu}\;, 
\end{eqnarray}
where $\eta_{ab}$ and $g_{\mu \nu}$ are, respectively, the internal flat space and spacetime metric, $\delta_{\nu}^{\mu}$ being the Kronecker delta function. The use of Greek and Roman indices distinguishes between curved spacetime and internal space, respectively.  In this framework, gravity can be described in terms of spin connection $A_{\mu}^{ab}$ and tetrad fields $e_{\mu}^a$. For the exploration of neutrino oscillation in curved spacetime, we adopt the methodology described in Ref.~\cite{Barick:2023wxx}. The spin connection can be written as the sum of two parts: one is purely universal gravitational part $(\omega_{\mu}^{ab})$ and the other is the non-universal part, called ``contorsion" $(\Lambda_{\mu}^{ab})$. Contorsion field is a non-dynamical field which can be expressed in terms of vector and axial current density of fermions. Thus, mathematically, spin connection can be expressed as,
\begin{eqnarray}
    A_{\mu}^{ab} = \omega_{\mu}^{ab} + \Lambda_{\mu}^{ab}\;.
    \label{spin}
\end{eqnarray}

$A_{\mu}^{ab}$ follows the tetrad postulate which is given by
\begin{equation}
    \qquad 	e^\lambda_a \partial_\mu e^a_\nu + 	A_{\mu}{}^{a}{}_{b} e^b_\nu e^\lambda_a - \Gamma^\lambda{}_{\mu \nu} = 0,
    \label{tetrad-postulate}
\end{equation}
where 
$\Gamma^\lambda{}_{\mu \nu}$
  is the Christoffel symbol. The field strength of the spin
connection can be written as,

\begin{align}\label{Ricci}
	F_{\mu\nu}^{ab} &= \partial_\mu A_\nu^{ab} - \partial_\nu A_\mu^{ab} + A_{\mu}{}^{a}{}_{c} A_{\nu}^{cb} -  A_{\nu}{}^{a}{}_{c} A_{\mu}^{cb}\,.
\end{align}
 The spin connection contributes to the covariant derivative ($D_{\mu}$) in the Lagrangian formulation for fermions. In presence of spin connection, the modified covariant derivative has the form,
\begin{equation}
    D_{\mu} \psi_i = \partial_{\mu} \psi_i - \frac{i}{4} A_{\mu}^{ab} \sigma_{ab} \psi_i  ~~~~~~ {\rm{where}}~~~~~ \sigma_{ab} = \frac{i}{2} [\gamma_{a},\gamma_{b}]_{-}
\end{equation}

The action for fermions ($\psi_i$) in presence of gravity can be written as \cite{Chakrabarty:2019cau,Barick:2023wxx}, 
\begin{equation}
    S= \int |e|~ d^4 x \left[ \frac{1}{2 \kappa} F_{\mu \nu}^{ab} e_{a}^{\mu} e_b^{\nu} + {\mathscr L}_\psi
    \right],
    \label{act}
    \end{equation}
where $|e|=\sqrt{g}$
  is the determinant of the tetrad, 
$\kappa=8 \pi G$ is the gravitational constant and ${\mathscr L}_\psi$ represents the fermionic Lagrangian which incorporates contributions from both torsion-free ($\omega_{\mu}^{ab}$) and torsional ($\Lambda_{\mu}^{ab}$) components of the spin connection,

\begin{align}
	{\mathscr L}_\psi &= \sum\limits_{i}	\frac{i}{2} \left(\bar{\psi}_i\gamma^\mu\partial_\mu\psi_i - 
	\partial_\mu\bar{\psi}_i\gamma^\mu\psi_i 	-\frac{i}{4} A_{\mu}^{ab}   \,
	\bar{\psi}_i[\sigma_{ab}, \gamma_c ]_{_+} \psi_i e^{\mu c} \right) \, +i m\bar\psi_i\psi_i\;. 
\end{align}
After optimizing $S$ with respect to $\Lambda_{\mu}^{ab}$, we obtain,
\begin{equation}
    \Lambda_{\mu}^{ab} = \frac{\kappa}{8} e_{\mu}^c \sum_i \bar{\psi}_i [\gamma_c,\sigma^{ab}]_{+} \psi_i\;.
    \label{a8}
\end{equation}

By applying the identity $[\gamma_c,\sigma_{ab}]_{+}=2 \epsilon_{abcd} \gamma^d \gamma^5$ in eq.~\ref{a8} and expressing $\psi$ in terms of its left-handed ($\psi_{L}$) and right-handed $(\psi_R)$ chiral components, the expression of $\Lambda_{\mu}^{ab}$ expands to,
\begin{equation}
    \Lambda_{\mu}^{ab} = \frac{\kappa}{4} \epsilon^{abcd} e_{c \mu} \sum_i \left(-\lambda_L^i \bar{\psi}_{iL} \gamma_d \psi_{iL} + \lambda_{R}^i \bar{\psi}_{iR} \gamma^d \psi_{iR} \right),
\end{equation}
where $\lambda_{L,R}^i$ are the coupling constants corresponding to the left-handed and right-handed chiral components of $\psi_i$.
 Substituting $\Lambda_{\mu}^{ab}$ back into the action, an effective four-fermion interaction emerges as \cite{Barick:2023wxx}

\begin{equation}\label{4fermi}
	-\frac{3\kappa}{16}\left(\sum\limits_i \left(-\lambda^i_{L}\bar{\psi}^i_{L} \gamma_a \psi^i_L + \lambda^i_{R}\bar{\psi}^i_{R} \gamma_a \psi^i_{R}\right)\right)^2\,.
\end{equation}

This interaction can also be rewritten in terms of vector and axial currents as
\begin{equation}\label{L-quartic}
	-\frac{1}{2}\left(\sum\limits_i \left(\lambda_i^V \bar{\psi}^i \gamma_a \psi^i + \lambda_i^A \bar{\psi}^i \gamma_a\gamma^5 \psi^i\right)\right)^2\,,
\end{equation}
where $\lambda^{V, A} = \frac{1}{2}(\lambda_R \pm \lambda_L)$\, and  a factor of $\sqrt{\frac{3\kappa}{8}}$\, is absorbed in $\lambda_{V, A}\,.$
The Dirac equation thus modified as
\begin{equation}\label{NLD}
	\gamma^\mu  \partial_\mu \psi^i - \frac{i}{4}\omega_{\mu}{}^{ab} \gamma^\mu \sigma_{ab} \psi^i
	+  m\psi^i
	+i\left(\sum_f\left(\lambda^f_{V}\bar{\psi}^f \gamma_a \psi^f + \lambda^f_A \bar{\psi}^f \gamma_a\gamma^5 \psi^f\right)\right)
	\left(\lambda^i_{V}\gamma_a \psi^i + \lambda^i_A \gamma_a\gamma^5 \psi^i\right) = 0\,,
\end{equation}
where $\psi^i$ is the fermion field whose equation we want, and $\psi^f$ represents  the field of any fermion,  the sum  running over all species.

In this analysis, we concentrate on the influence of gravity on neutrino mass eigenstates, treating all fermions other than neutrinos as a background. 
In eq. \ref{L-quartic}, upon decomposing the gamma matrices, we encounter four distinct elements: the temporal and spatial components of vector currents, axial charges, and the spatial components of axial currents. Among these components, only the temporal component of the vector currents remains relevant, effectively representing the background matter density with $f=e,p,n$ (i.e., electrons, protons, and neutrons, respectively) expressed as $(\sum\limits_{f=e,p,n}\lambda_{fV} n_f )$, where $n_f= \langle \overline{\psi}_f \gamma^0 \psi_f \rangle$. For further details, we refer to Ref.~\cite{Barick:2023wxx,Chakrabarty:2019cau}. This extra term contributes an additional component to the Hamiltonian, influencing the mass of the neutrinos.

\bibliography{main}

\end{document}